\documentclass[letterpaper]{article}
\usepackage[draft]{aaai2026}
\usepackage{times}
\usepackage{helvet}
\usepackage{courier}
\usepackage[hyphens]{url}
\usepackage{graphicx}
\usepackage{longtable}
\usepackage[hyphens]{url}
\usepackage{array}

\usepackage[numbers,square]{natbib}
\usepackage{caption}
\usepackage{booktabs}
\usepackage{longtable}
\usepackage{array}
\usepackage{calc}
\usepackage{makecell}
\usepackage{textcomp}
\providecommand{\tightlist}{\setlength{\itemsep}{0pt}\setlength{\parskip}{0pt}}

\newcommand{\keywords}[1]{\noindent\textbf{Keywords:} #1}

\title{\#MakeBeefGreatAgain: A Cross-Platform Analysis of Early \#MAHA Discourse}
\author{Haoning Xue\textsuperscript{1,2,*}, Yue Li\textsuperscript{3}, Benjamin A. Lyons\textsuperscript{1,2}, and Andy J. King\textsuperscript{1,2}}
\affiliations{\textsuperscript{1}Department of Communication, University of Utah, USA\\
\textsuperscript{2}Cancer Control and Population Sciences, Huntsman Cancer Institute, University of Utah, USA\\
\textsuperscript{3}Department of Communication Studies, University of Kansas, USA\\
*Contact corresponding author: \texttt{haoning.xue@utah.edu}}
\begin{document}
\maketitle

\begin{abstract}
Make America Healthy Again (MAHA) is a health-related campaign slogan proposed by Robert F. Kennedy Jr. and later incorporated into President Trump’s political coalition. While \#MAHA quickly circulated beyond the campaign itself and became a prominent hashtag for public discussion, it remains unclear whether this public discourse reflected, reshaped, or diverged from the campaign’s stated agenda. This study presents a large-scale, cross-platform analysis of early \#MAHA public discourse between September 2024 and January 2025, using the framework of Agenda-Melding Theory. Drawing on 41,819 \#MAHA-related posts, this study combines structural topic modeling, interrupted time-series analysis, and AI-assisted data annotation to examine the thematic structure and temporal dynamics. The most prominent finding is the substantial disconnect between \#MAHA public discourse and the campaign’s stated MAHA agenda: 81.3\% of posts did not engage any of the campaign’s five campaign priorities. There were also pronounced cross-platform differences, with online platforms clustering into three broad discourse environments: (a) grassroots partisan-support spaces, (b) informational sources, and (c) health-focused spaces. \#MAHA functioned less as a unified campaign agenda than as a symbolic frame interpreted differently across platforms. More broadly, this study provides useful empirical insight into how campaign slogans are reinterpreted and how public agendas are formed, amplified, and transformed in today’s fragmented digital environments.
\end{abstract}




\section{Introduction}\label{introduction}

Make America Healthy Again (MAHA) is a campaign slogan proposed by Robert F. Kennedy Jr., which was later incorporated into President Trump’s political coalition. This campaign highlights five issue priorities: tackle the chronic disease epidemic, advance regenerative agriculture, protect natural habitats, expose corporate corruption, and eliminate environmental toxins \citep{maha_make_2025}. After the new administration took office, the MAHA agenda began to take shape in policy, including an MAHA-related executive order in February 2025 \citep{the_white_house_establishing_2025}, followed by legislation introduced in 38 states \citep{tschopp_state_nodate}. While the MAHA campaign centers on health issues, it is essentially health-political as it emerged within an active campaign context.

Alongside these legislative developments, \#MAHA quickly circulated beyond the campaign itself and became a prominent hashtag for public discussion across various online platforms, appearing in over 300,000 TikTok videos \citep{tiktok_maha_2026} and 600,000 Instagram posts \citep{instagram_maha_2026} alone. The \#MAHA discourse has been actively shaped by key figures in the campaign, such as Robert F. Kennedy Jr. and Casey Means \citep{hooper_maha_2026}, often through content centered on fitness, food, and other lifestyle practices associated with the MAHA brand \citep{howe_whole_2025}. Rooted partly in the broader Make America Great Again (MAGA) political identity \citep{sable-smith_watch_2026}, \#MAHA has extended beyond campaign elites and high-profile influencers \citep{reed_rfk_2025}. For example, communities like ``MAHA moms'' are becoming active participants with significant online presence \citep{reed_rfk_2025}. \#MAHA has become a broader symbolic frame through which diverse publics make sense of and engage with this politically health-related issue.

This raises an important question: How did the \#MAHA public discourse reflect, (re)shape, or diverge from the MAHA campaign's stated priorities? Campaign agendas do not simply replicate themselves through public discourse. The Agenda-Melding Theory (AMT) \citep{shaw_individuals_1999} suggests that individuals actively meld multiple agendas--from campaigns, news media, and social groups--resulting in some issues being amplified, others suppressed, and new issue agendas emerging. From this perspective, early \#MAHA public discourse may not have fully centered on the campaign's proposed priorities. Examining this alignment retrospectively is important because it reveals how campaign priorities are interpreted by the public and circulated in the broader information environment.

This question is particularly important during the early diffusion of a rapidly evolving and politically oriented health movement. In its early stages, the meaning of a campaign slogan is still being actively constructed across online platforms. In particular, the 2024 presidential election provides an important context for understanding this process, as election-related dynamics likely shaped the content and volume of \#MAHA discourse over time. Therefore, this study focuses on the early stages precisely to capture the process of public discourse formation before its meaning and engagement patterns stabilize.

Public discourse is also not uniform in the online space. Different spaces and platforms represent distinct online ecosystems, shaped by their respective audiences, affordances, and norms \citep{alipour_cross-platform_2024, li_platform_2026, sude_different_2023}. Prior research suggests that online spaces are ideologically fragmented, such that platforms differ not only in information exposure but also in community identities \citep{di_martino_ideological_2025}. For example, people usually turn to X (formerly Twitter) for informational purposes, whereas Instagram is often used for social interaction and entertainment \citep{pelletier_one_2020}. As people tend to select platforms and communities that fit their identities and interests \citep{slater_reinforcing_2007}, \#MAHA discourse may vary substantially across platforms. However, it remains unclear which platforms gave the most attention to \#MAHA and how different platforms highlighted different aspects of the campaign agenda.

To address these questions, this study analyzes \#MAHA-related posts (N = 41,819) collected from September 2024 (early use by RFK, Jr.) to January 2025 (the start of President Trump's second term) across seven online sources, including six social media platforms (i.e., TikTok, Instagram, Facebook, X, Reddit, and YouTube) and web articles. Using both structural topic modeling and time-series analysis, we examine how \#MAHA discourse evolved in the campaign's early stage, (mis)aligned with the campaign agenda, and varied across online platforms. This study provides broader insight into how politically health-related narratives evolve within the fragmented online information environment. Specifically, we ask three research questions below.

\begin{quote}
\textbf{RQ1.} How did \#MAHA public discourse evolve over time during the early stage of the introduction of, and integration of, MAHA into the presidential campaign?

\textbf{RQ2.} To what extent did \#MAHA public discourse align with the campaign's proposed MAHA agenda?

\textbf{RQ3.} How did \#MAHA public discourse vary across online platforms?
\end{quote}

\section{Results}\label{results}

\subsection{Early \#MAHA Discourse}\label{early-maha-discourse}

Using Structural Topic Modeling (STM), four dominant themes emerged in early \#MAHA discourse: (1) healthy lifestyle, (2) health policy, (3) electoral campaign, and (4) patriotic partisan support. Table 1 summarizes the four topics and provides representative example posts.

Topic 1 (n = 8,978; 21.6\%) centers on healthy lifestyle discussions. Posts in this topic emphasized nutrition and fitness, including home cooking, supplements, exercise, and other everyday wellness practices. This content is often framed through natural living and homestead-oriented values.

Topic 2 (n = 6,766; 16.2\%) focused more on policy-oriented health discussions. Posts associated with this topic highlighted systemic health issues in the U.S. and proposed reforms to the food, and healthcare regulatory systems. Common topics included chronic disease, processed foods, and healthcare access.

Topic 3 (n = 9,669; 23.2\%) shifted away from substantive health discussion and toward campaign-related political messaging. Posts in this topic centered on endorsements, appointments, and event promotion. The \#MAHA hashtag here was repeatedly used as a slogan rather than a substantive discussion of health issues.

Topic 4 (n = 16,246; 39.0\%) was the largest topic and consists primarily of grassroots expressions of support and celebration. As in Topic 3, \#MAHA was used primarily as a slogan rather than as a framework for discussing specific health issues. However, whereas Topic 3 was more closely tied to campaign events and political figures mentioned more often in web articles and YouTube videos, Topic 4 was dominated by ordinary users expressing enthusiasm, solidarity, and patriotic support concentrated on TikTok. 

\begin{table}[t]
\centering
\small
\begin{tabular}{p{0.06\textwidth} p{0.06\textwidth} p{0.06\textwidth} p{0.18\textwidth} p{0.20\textwidth} p{0.28\textwidth}}
\hline
\textbf{Topic} & \textbf{M (SD)} & \textbf{n (\%)} & \textbf{Highest-probability keywords} & \textbf{FREX keywords} & \textbf{Example posts} \\
\hline

Health lifestyle &
0.22 (0.26) &
8978 (21.6\%) &
maha, make, food, healthy, makeamericahealthyagain, health, donald, day, well, come &
diet, body, fit, carnivore, ingredient, mom, heal, fat, mind, beef &
``Want to Make America Healthy Again? Here's a lifestyle blueprint you can try...''

``RAW MEAT SO NATRUL MAKE BIG MUSCLE FOR MERICA MAHA'' \\
\hline

Health policy &
0.19 (0.21) &
6766 (16.2\%) &
health, rfk, food, american, america, administration, public, new, agency, medical &
medical, second, fluoride, drug, push, propose, remove, medicare, sweep, medicaid &
``Fluoride in water has slashed tooth decay by 25\%, but could MAHA's push for change disrupt decades of proven public health benefits?''

``If confirmed, RFK Jr. will take the reins at a \$1.7 trillion agency that oversees vaccines, medicines, scientific research, public health infrastructure, and the health care plans of millions of Americans.'' \\
\hline

Electoral campaign &
0.24 (0.26) &
9669 (23.2\%) &
trump, kennedy, donald, make, president, robert, healthy, health, america, election &
president, bhattacharya, jay, thursday, launch, pac, pledge, chosen, october, colorado &
``BREAKING: RFK Jr to be the Secretary of Health and Human Services! This is HUGE! MAKE AMERICA HEALTHY AGAIN!''

``Kansas senator announces creation of Senate's `Make America Healthy Again' caucus'' \\
\hline

Patriotic partisan support &
0.35 (0.36) &
16,246 (39.0\%) &
trump, maha, maga, america, vote, republican, usa, rfkjr, elect, democrat &
maga, usa, donaldtrump, americafirst, trumpvance, patriot, trumpsupport, elonmusk, aff, jdvance &
``MAGA + MAHA = WIN!''

``WE ARE UNITED IN FIGHTING FOR AMERICA LET'S GO \#Trump2024 \#MAGA \#MAHA'' \\
\hline
\end{tabular}
\caption{Structural topic modeling results for early \#MAHA public discourse, including (a) the mean and standard deviation of STM-derived topic prevalence, (b) the number and percentage of posts for which each topic was the dominant topic, (c) highest-probability keywords (most frequent within the topic), (d) FREX keywords (high-frequency terms that are exclusive to the topic), and (e) representative example posts.}
\label{tab:table_1}
\end{table}

\subsection{Temporal Dynamics of \#MAHA Discourse}\label{temporal-dynamics-of-maha-discourse}

After identifying the four major themes in early \#MAHA discourse, we examined how these themes changed over time and the role of the 2024 presidential election in this process to answer RQ1. We conducted interrupted time series analyses to identify the temporal dynamics of early \#MAHA discourse. See Figures 1-2 for the visualization of temporal patterns and Tables S1-S2 in the Supplementary Information (SI) for full model results. The overall \#MAHA post volume showed a pronounced spike on the election day (Figure 1 Panel A). Daily post volume increased significantly on the election day and then declined significantly during the post-election period. This pattern suggests that overall public attention to \#MAHA peaked around the election but was not sustained in the short-term afterward.

However, this overall pattern presented differently across topics (Figure 2). First, Topic 1 (healthy lifestyle) showed a significant post-election increase, despite no significant overall time trend or immediate election-day effect. This suggests that \#MAHA public discourse was more oriented toward health lifestyle and wellness practices after the election. Second, while there were no significant election-day or post-election changes, Topic 2 (health policy) showed a steady, significant positive time trend. It suggests that there was a gradual increase in public attention to health policy and related issues over time, largely independent of the election itself. By contrast, Topics 3 and 4 were more closely connected to the election context. Topic 3 (electoral campaign) showed a significant increase on the U.S. election day. Topic 4 (patriotic partisan support) showed a significant post-election decline, indicating that this type of grassroots partisan expression became less prevalent after the election and remained lower thereafter.

Overall, these findings suggest that although early \#MAHA discourse revolved heavily around the presidential election, various themes showed distinct temporal trends. While health policy remained a gradually increasing topic, post-election \#MAHA discourse became more focused on healthy lifestyle content. However, the discussion on the election, campaign, and candidacy with \#MAHA as a major slogan hashtag was more concentrated on the electoral moment and sustained afterward. 

\begin{figure}[htb]
\centering
\includegraphics[width=0.8\columnwidth]{fig_1.png}
\caption{Interrupted time-series analysis results of (A) daily \#MAHA post count and (B) daily proportion of MAHA campaign-relevant posts from Sep 05, 2024 to Jan 10, 2025. Gray points indicate observed daily values. The vertical dashed line marks the presidential election day (i.e., November 5, 2024) as the interruption point. Colored solid lines represent fitted interrupted time-series trends with 95\% CIs. The gray dashed lines represent the expected pattern in the absence of an election-related interruption (i.e., the counterfactual post-election trends estimated from the pre-election trajectory).}
\label{fig1}
\end{figure}

\subsection{\#MAHA Public Discourse Diverged from MAHA Campaign Agenda}\label{maha-public-discourse-diverged-from-maha-campaign-agenda}

To answer RQ2, we examined how the STM-derived \#MAHA topics aligned with the five campaign-stated priorities (see Figure 3 for visualization and Table S3 for contingency table). Overall, most \#MAHA posts did not map onto any of the campaign's five stated priorities: 81.3\% of posts (n = 34,000) were classified as not related to any campaign priority, whereas only 18.7\% (n = 7,659) aligned with one of the five priorities tied to the stated campaign agenda. Among agenda-related posts, Agenda Priority 1 (Chronic Disease Epidemic) accounted for the largest share (n = 4,708, 61.5\%), whereas Agenda Priority 3 (Habitat Preservation) received the least public attention (n = 6, 0.1\%).

A Pearson chi-square test further showed a significant association between STM-derived topic and campaign agenda classification ($\chi^2$ = 10,484, p \textless{} .001; Cramer's V = 0.29). This means that campaign agendas were distributed unevenly across the four topics. Therefore, we examined standardized residuals to identify specific patterns of discourse-campaign alignment and misalignment (Table S4).

\begin{figure}[htb]
\centering
\includegraphics[width=0.8\columnwidth]{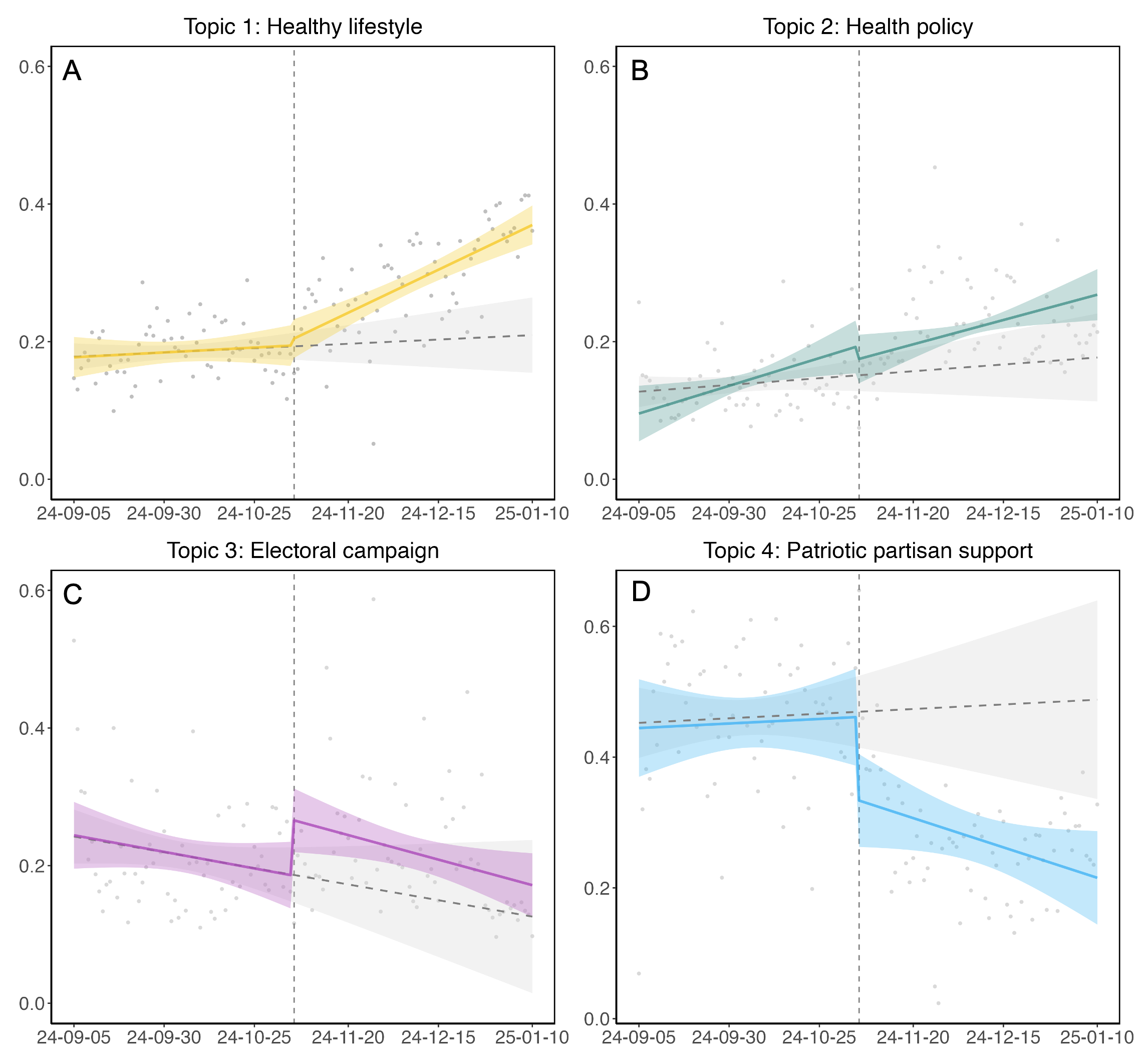}
\caption{Interrupted time-series analysis results of STM-derived topic prevalence from Sep 05, 2024 to Jan 10, 2025. Gray points indicate observed daily values. The vertical dashed line marks the presidential election day (i.e., November 5, 2024) as the interruption point. Colored solid lines represent fitted interrupted time-series trends with 95\% CIs. The gray dashed lines represent the expected pattern in the absence of an election-related interruption (i.e., the counterfactual post-election trends estimated from the pre-election trajectory).}
\label{fig2}
\end{figure}

The residual patterns showed that Topic 1 (healthy lifestyle) and Topic 2 (health policy) were the main topics in which campaign priorities appeared. Posts aligned with Agenda Priority 1 (Chronic Disease Epidemic), 2 (Regenerative Agriculture), and 5 (Removing Toxins from the Environment) were more common in these two topics. This pattern was consistent with the substantive focus of the thematic topics: Topic 1 centered on everyday health practices and lifestyle choices, whereas Topic 2 focused more on policy, regulation, and systemic reform. Further, Agenda Priority 4 (Combatting Corporate Corruption) appeared especially often in Topic 2, suggesting that corruption-related campaign priorities were discussed mainly in policy-oriented posts. By contrast, Topic 3 (electoral campaign) and Topic 4 (patriotic partisan support) showed very little alignment with the campaign’s stated agenda priorities. Rather than engaging with specific issue priorities, these topics were dominated by campaign-centered messaging and patriotic expressions. Overall, much of the election-oriented circulation of \#MAHA did not translate into discussion of the campaign’s concrete agenda priorities.

\begin{figure}[htb]
\centering
\includegraphics[width=0.9\columnwidth]{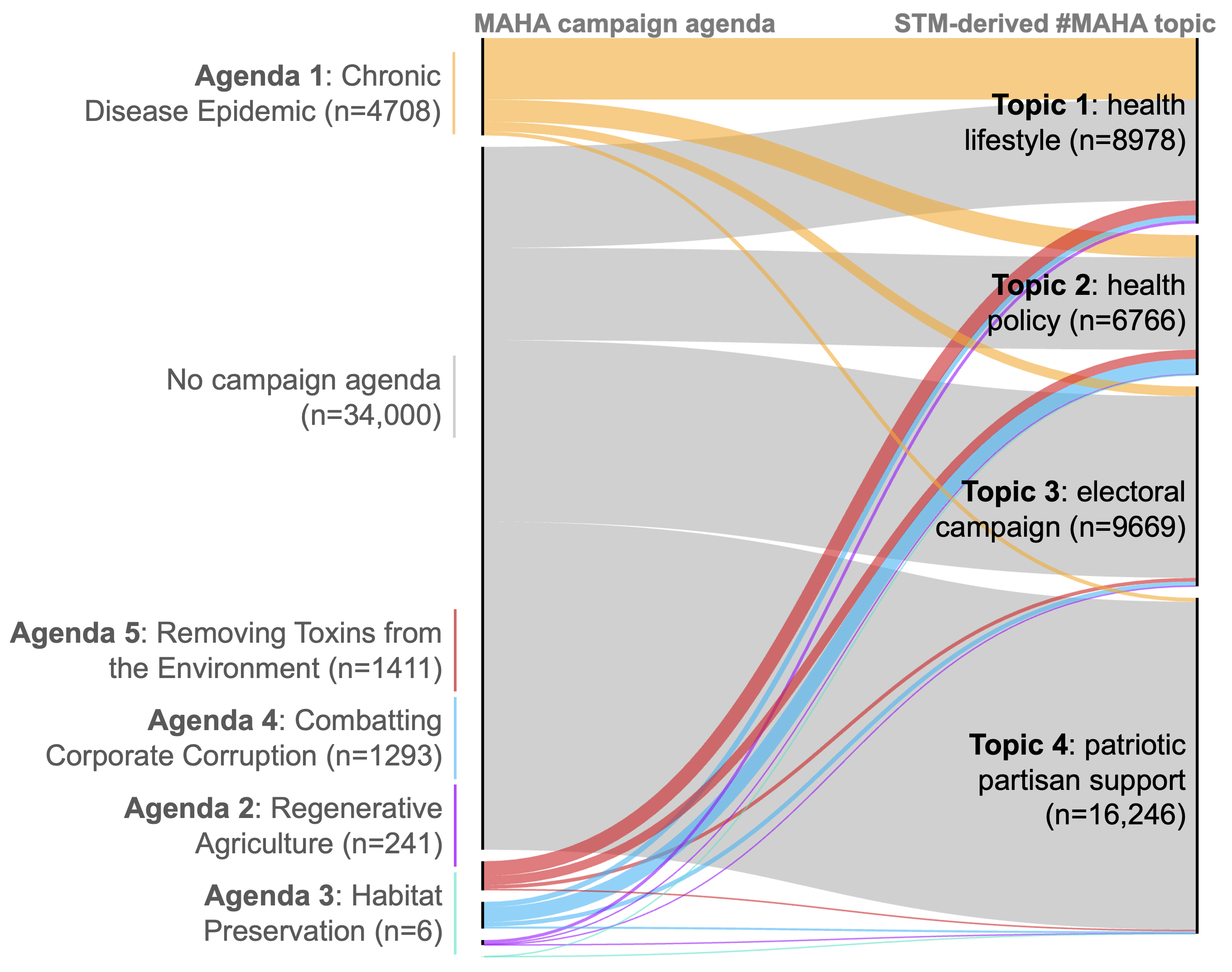}
\caption{Distribution of MAHA campaign agenda categories across STM-derived \#MAHA topics in a Sankey diagram.}
\label{fig3}
\end{figure}

\subsection{Early \#MAHA Public Discourse Across Seven Online Platforms}\label{early-maha-public-discourse-across-seven-online-platforms}

To answer RQ3, we examined how the prevalence of STM-derived topics varied across online platforms using a Pearson chi-square test (see Figure 4 for visualization and Tables S5-S6 for contingency table and standardized residuals). We found a relatively strong association between STM-derived topics and online platforms ($\chi^2$ = 30,414, p \textless{} .001; Cramer's V = 0.49). This result indicates that different online platforms emphasized markedly different aspects of \#MAHA discourse.

Most notably, TikTok stood out as a distinct platform, heavily concentrated on Topic 4 (patriotic partisan support) and, to a lesser extent, Topic 1 (healthy lifestyle). It suggests that \#MAHA primarily functioned as a symbolic frame for grassroots partisan expression and lifestyle-oriented health discussions. On the contrary, web articles, X, YouTube, and Facebook represented a more informational aspect of \#MAHA discourse, leaning toward Topic 2 (health policy) and Topic 3 (electoral campaign). Lastly, Instagram and Reddit reflected a more health-focused discussion overall, with a greater emphasis on Topics 1 and 2. Overall, these patterns suggest that \#MAHA did not circulate as a uniform discourse across online platforms. Instead, different platforms amplified different aspects of the \#MAHA discourse, forming three broad clusters: (a) grassroots partisan support space, (b) informational sources, and (c) health-oriented spaces.

\section{Discussion}\label{discussion}

Combining structural topic modeling, interrupted time-series analysis, and AI-assisted data annotation, this study provides a large-scale, cross-platform examination of early public discourse surrounding \#MAHA through the lens of Agenda-Melding Theory (AMT) \citep{shaw_individuals_1999}. The most prominent finding is the substantial disconnect between \#MAHA public discourse and the MAHA campaign's stated agenda: 81.3\% of posts did not engage any of the campaign's five campaign agenda priorities. Early \#MAHA discourse was strongly tied to the presidential election in general, with campaign-related and partisan-support content dominating the early period. After the election, \#MAHA discourse became increasingly oriented toward healthy lifestyle and health policy discussions. We also observed substantial cross-platform differences in their differentiated focuses on grassroots patriotic partisan expressions, informational content, or health discussions.

\begin{figure}[htb]
\centering
\includegraphics[width=0.9\columnwidth]{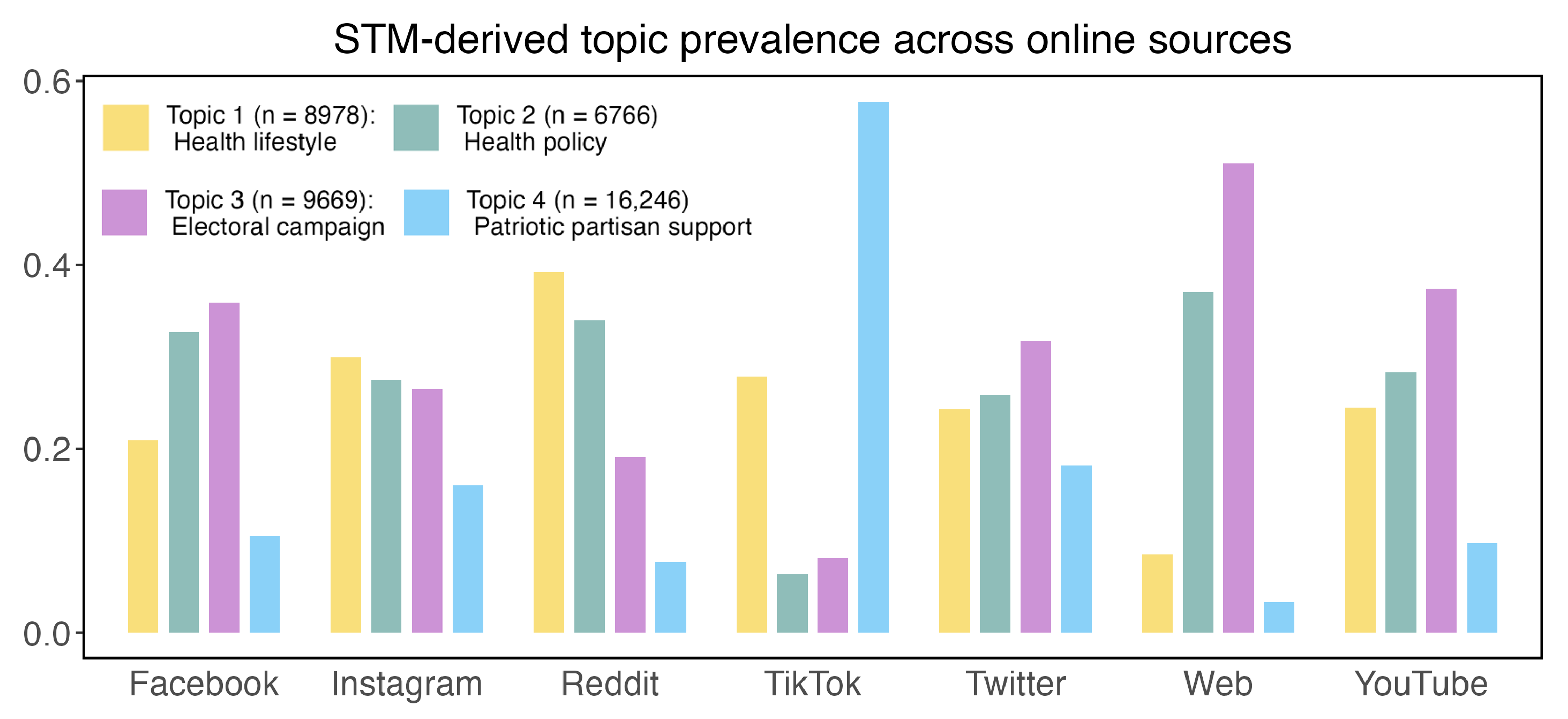}
\caption{STM-derived topic prevalence across seven online platforms.}
\label{fig4}
\end{figure}

The disconnect between \#MAHA public discourse and the MAHA campaign agenda priorities is consistent with AMT \citep{shaw_individuals_1999}. Rather than passively reproducing elite agendas, AMT suggests that individuals actively reinterpret, prioritize, and downplay issues from campaigns, media, and social groups. Early \#MAHA public discourse did not simply echo the campaign's stated priorities. Instead, it selectively amplified lifestyle- and policy-oriented health discussions while giving far less attention to other campaign agendas, such as habitat preservation and regenerative agriculture. This process matters because public-melded agendas may shape which issues become visible, socially resonant, and influential in the broader information environment. Prior research suggests that public discourse on social media can affect individual behavior \citep{vargo_public_2022}, influence the agenda priorities of traditional media \citep{luo_how_2019}, and interact with both social and traditional media in iterative agenda-setting processes \citep{gilardi_social_2022}. In the context of the presidential election, our study demonstrates how a campaign slogan can circulate widely in public discourse while becoming partially detached from the campaign agenda it was originally meant to represent.

This study also extends the cross-platform implications of AMT by showing that different online platforms participated in \#MAHA discourse in systematically different ways. TikTok was dominated by grassroots patriotic and partisan support, whereas web articles, X, YouTube, and Facebook were more oriented toward informational and campaign-related content. Instagram and Reddit were relatively more health-focused. These patterns speak to an increasingly fragmented information environment \citep{steppat_selective_2022} in which fragmentation is not only about exposure to different issues but also about the emergence of different perspectives on the same issue across platforms \citep{porten-chee_fragmentation_2019}. This is consistent with prior research that platforms function as distinct communicative environments, driven by different affordances, norms, and cultures \citep{li_platform_2026, sude_different_2023}. In turn, these platform-specific characteristics can shape which aspects of an issue become most visible and how it is interpreted in public discourse.

Meanwhile, this process is not driven solely by platform structure. The uses and gratifications theory suggests that audiences do not passively absorb whatever discourse a platform presents; they actively select media environments that fit their interests, identities, and communicative goals \citep{wood_active_2014, krcmar_uses_2009}. Therefore, platform differences in \#MAHA discourse likely reflect both platform-specific communicative environments and users' active role in choosing, reproducing, and amplifying aspects of the discourse that resonate with their interests and identities. This point also extends AMT in the sense that agenda melding occurs not only through selective engagement with campaign, media, and social agendas, but also through active participation in platform-specific environments where those agendas are reinterpreted and reshaped.

Overall, this large-scale cross-platform analysis shows how public discourse can diverge from elite agendas and take on different perspectives across online platforms. \#MAHA emerged as a flexible public discourse shaped by audience selection, platform-specific environments, and evolving political context. By tracing and analyzing these dynamics, this study provides useful empirical insight into how campaign slogans are reinterpreted and how public agendas are formed, amplified, and transformed in today's fragmented digital environments.

\section{Materials and Methods}\label{materials-and-methods}

\subsection{Data Collection}\label{data-collection}

\begin{figure}[htb]
\centering
\includegraphics[width=0.9\columnwidth]{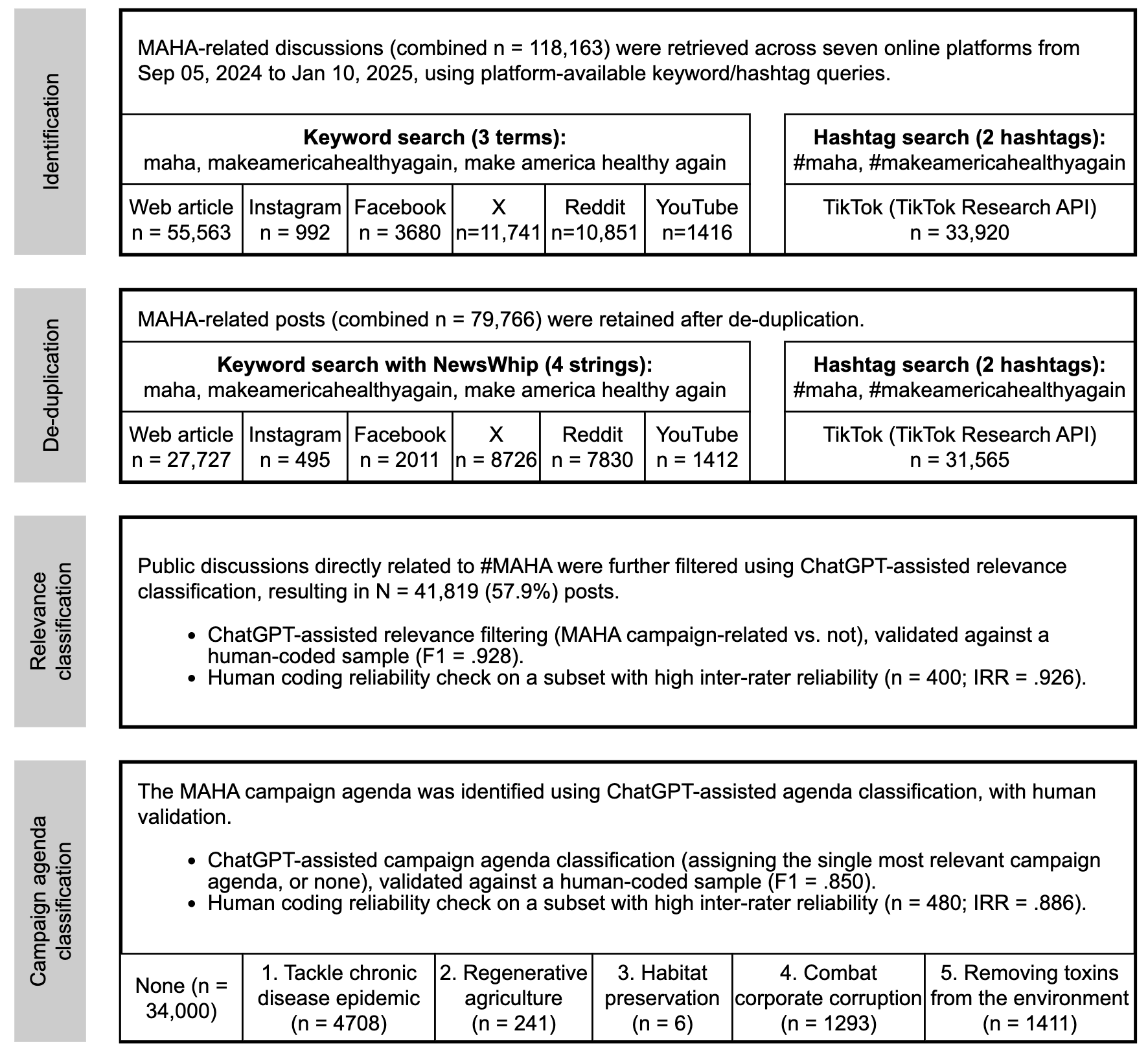}
\caption{Flow diagram of data collection, ChatGPT-assisted relevance classification, and ChatGPT-assisted campaign agenda classification.}
\label{fig5}
\end{figure}

This study draws on a cross-platform dataset of N = 41,819 posts related to \#MAHA across seven online sources, including six social media platforms (i.e., TikTok, Instagram, Facebook, X, Reddit, and YouTube) and web articles (see details below). Posts from these sources were connected between September 05, 2024 (early use by RFK, Jr.) and January 10, 2025 (the start of President Trump's second term). This timeframe was chosen to capture the early \#MAHA discourse, spanning approximately two months before and after the November 5, 2024, U.S. presidential election. See Figure 5 for details on data collection, data processing, and AI-assisted agenda annotation. The data and replication code for this study are publicly available on OSF: \url{https://osf.io/mb5kf}.

To construct this cross-platform dataset, we adopted a three-step data collection approach. First, we used the TikTok Research API \citep{tiktok_research_2024} to collect video descriptions and associated metadata using two \#MAHA-related hashtags (see Figure 5 for the list of search terms). Second, we used the YouTube Data API \citep{youtube_data_api_api_2025} to collect video metadata using three \#MAHA-related terms (see Figure 5 for the full list). Lastly, we used NewsWhip Spike \citep{newswhip_newswhip_2026-1}, a commercial social listening agency, to collect \#MAHA-related content from the remaining five platforms, including web articles, Instagram, Facebook, X, and Reddit.

NewsWhip Spike is a commercial analytics platform that tracks and ranks online news and social media content from nearly 500,000 websites using real-time engagement metrics (e.g., shares, reactions, and comments) \citep{newswhip_real-time_2026}. It aggregates publicly available content from multiple platforms to identify stories receiving high levels of attention, providing a large and diverse cross-platform trending content for analysis. However, because NewsWhip Spike prioritizes engagement metrics, the resulting dataset reflects content that has achieved substantial visibility and interaction rather than the full universe of online content. As such, platform comparisons using this data reflect comparisons of high-visibility content rather than overall discourse. Since hashtags are not used consistently across platforms, this stage relied on a keyword-based search using the same \#MAHA-related terms.

Across all collection steps, the initial dataset contained n = 118,163 posts. After removing duplicate entries based on URL, the dataset was reduced to n = 79,766 unique posts. For web articles, this step retrieved n = 11,855 web articles from 2,240 domains. These sources included both general web portals (e.g., yahoo.com, aol.com) and news websites (e.g., ussanews.com, foxnews.com). See Table S7 for summary statistics across platforms.

\subsection{Relevance Classification}\label{relevance-classification}

Following data collection, we conducted a relevance-filtering procedure to remove posts not directly related to the MAHA campaign. This step was necessary because ``MAHA'' can carry unrelated meanings in other contexts and languages, including as a Hindi prefix meaning ``great'' and as a common Arabic female name. To identify relevant posts, two trained researchers first manually annotated a random subset of n = 400 posts for relevance, achieving high inter-rater reliability (Cohen's kappa = .926). Second, we used ChatGPT-5.1 for relevance classification of the full dataset (see Table S8 for the complete instruction text). Compared with the human-coded benchmark, the model achieved high accuracy (F1 = .928). After relevance filtering, the final analytic sample consisted of N = 41,819 posts, representing 57.9\% of the original dataset.

\subsection{Campaign Agenda Annotation}\label{campaign-agenda-annotation}

We assessed how \#MAHA posts aligned with the five campaign agendas listed on the maha.vote website as of March 2025 \citep{maha_make_2025}. Although the website has been updated over time, this version served as the reference point for our coding scheme. Specifically, we applied ChatGPT-5.1 to annotate whether each post reflected the single most prominent campaign agenda, if any. The annotation instruction included the full campaign agendas and explanations for each agenda priority (see Table S8 for the complete instruction text).

Compared with the human-coded benchmark of a random subset of n = 480 posts by two trainer researchers (Cohen's kappa = .886), the model achieved high accuracy (F1 = .850). We found that n = 34,000 (81.3\%) posts were not related to any campaign agenda priority. To further validate this finding, two trained researchers reviewed a random subset of 340 posts classified as ``no agenda'' by GPT 5.1. This validation again showed high inter-rater reliability (Cohen's kappa = .879) and confirmed that 94.7\% of these posts (n = 322) were indeed unrelated to any campaign agenda.

\subsection{Structural Topic Modeling}\label{structural-topic-modeling}

We used \emph{stm} R package \citep{roberts_stm_2019} to identify thematic topics in this dataset. We selected structural topic modeling because it allows topic prevalence to be modeled as a function of document-level covariates, making it appropriate for examining how discourse changed over time. Before model estimation, we conducted both standard and context-specific text preprocessing. This included removing URLs, non-English words, generic stop words, and context-specific stop words (e.g., foryoupage, fyp). We conducted a K-search to select the optimal number of topics and fitted the final STM with topic prevalence as a function of time, allowing topic proportions to vary over time. We identified and interpreted 4 topics using standard STM diagnostics (e.g., highest-probability and FREX terms; see Table 1 for details).

\subsection{Interrupted Time Series Analysis}\label{interrupted-time-series-analysis}

To investigate how the 2024 presidential election plays a role in shaping \#MAHA discourse, we conducted interrupted time series analysis using the \emph{nlme} R package \citep{pinheiro_nlme_2026}. We first aggregated the data to the daily level, calculating the total number of posts per day and the mean STM-derived topic prevalence for each of the four topics. To account for temporal autocorrelation in daily time series, we estimated generalized least squares (GLS) models with autoregressive-moving-average (ARMA) correlation structures. The optimal parameters were selected based on the Akaike Information Criterion (AIC). Then we refitted the final models using daily post count and daily topic prevalence as dependent variables. Each model included three independent variables: a continuous time trend, an indicator for the election day interruption, and a post-election time trend. We estimated five models in total.

\subsection{Limitations}\label{limitations}

This study is not without limitations. First, the analysis was based primarily on textual content, including post texts, video descriptions, and brief article summaries. Therefore, the multimodal content in \#MAHA discourse was not fully captured. In addition, non-English content and linked material (e.g., URLs) were not analyzed in detail. Second, content collected through NewsWhip may reflect biases in platform coverage because the service emphasizes trending content rather than a complete census of all posts. This limitation is especially relevant when comparing platforms with uneven sample sizes or different recommendation systems. Lastly, although interrupted time-series analysis helps identify temporal shifts around the presidential election, the study remains observational and correlational. The results support inference about temporal association, but not the causal effects of the election on discourse change.

\newpage
\section{References}\label{references}

\renewcommand{\refname}{}
\vspace{-2em}
\bibliography{bibliography}

\clearpage

\setcounter{secnumdepth}{0}

\newcolumntype{L}[1]{>{\raggedright\arraybackslash}p{#1}}
\newcolumntype{C}[1]{>{\centering\arraybackslash}p{#1}}
\renewcommand{\arraystretch}{1.16}
\setlength{\tabcolsep}{2pt}
\providecommand{\tightlist}{\setlength{\itemsep}{0pt}\setlength{\parskip}{0pt}}

\section*{Supplementary Information for\\
\#MakeBeefGreatAgain: A Cross-Platform Analysis of Early \#MAHA Discourse}

\vspace{5em}

\setcounter{table}{0}
\renewcommand{\thetable}{S\arabic{table}}

\setcounter{figure}{0}
\renewcommand{\thefigure}{S\arabic{figure}}

\begin{table}[htbp]
\centering
\begin{tabular}{L{0.12\textwidth} C{0.20\textwidth} C{0.09\textwidth} C{0.20\textwidth} C{0.09\textwidth}}
\toprule
 & \multicolumn{2}{c}{\textbf{Post count}} & \multicolumn{2}{c}{\textbf{Agenda-aligned post \%}} \\
\midrule
 & \textbf{b (SE)} & \textbf{p} & \textbf{b (SE)} & \textbf{p} \\
\midrule
Intercept & 47.96 (129.64) & .712 & 0.131 (0.03) & .000 \\
Time & 5.09 (3.55) & .154 & 0.001 (0.00) & .404 \\
Event & 647.15 (162.01) & .000 & -0.010 (0.03) & .756 \\
Post Event & -21.69 (4.93) & .000 & 0.002 (0.00) & .042 \\
\bottomrule
\end{tabular}
\caption{Interrupted time-series regression models of (A) daily \#MAHA post volume and (B) daily proportion of MAHA campaign-relevant posts around the 2024 U.S. presidential election.}
\label{tab:s1}
\end{table}

\begin{table}[htbp]
\centering

\begin{tabular}{L{0.1\textwidth} C{0.10\textwidth} C{0.06\textwidth} C{0.12\textwidth} C{0.06\textwidth} C{0.12\textwidth} C{0.06\textwidth} C{0.12\textwidth} C{0.06\textwidth}}
\toprule
 & \multicolumn{2}{c}{\textbf{Topic 1}} & \multicolumn{2}{c}{\textbf{Topic 2}} & \multicolumn{2}{c}{\textbf{Topic 3}} & \multicolumn{2}{c}{\textbf{Topic 4}} \\
 & \multicolumn{2}{c}{\textbf{(healthy lifestyle)}} & \multicolumn{2}{c}{\textbf{(health policy)}} & \multicolumn{2}{c}{\textbf{(electoral campaign)}} & \multicolumn{2}{c}{\textbf{(patriotic partisan support)}} \\
\midrule
 & \textbf{b (SE)} & \textbf{p} & \textbf{b (SE)} & \textbf{p} & \textbf{b (SE)} & \textbf{p} & \textbf{b (SE)} & \textbf{p} \\
\midrule
Intercept & 0.18 (0.02) & .000 & 0.09 (0.02) & .000 & 0.25 (0.03) & .000 & 0.44 (0.04) & .000 \\
Time & 0.000 (0.00) & .514 & 0.002 (0.00) & .014 & -0.001 (0.00) & .177 & 0.000 (0.00) & .798 \\
Event & 0.008 (0.02) & .711 & -0.019 (0.03) & .568 & 0.081 (0.03) & .020 & -0.126 (0.05) & .016 \\
Post Event & 0.002 (0.00) & .000 & -0.000 (0.00) & .804 & -0.000 (0.00) & .625 & -0.002 (0.00) & .161 \\
\bottomrule
\end{tabular}
\caption{Interrupted time-series regression models of daily STM-derived topic prevalence around the 2024 U.S. presidential election.}
\label{tab:s2}
\end{table}

\newpage
\begin{table}[htbp]
\centering
\begin{tabular}{L{0.06\textwidth} C{0.10\textwidth} C{0.15\textwidth} C{0.12\textwidth} C{0.12\textwidth} C{0.15\textwidth} C{0.21\textwidth}}
\toprule
 & \textbf{No campaign agenda} & \textbf{Agenda 1 (Chronic Disease Epidemic)} & \textbf{Agenda 2 (Regenerative Agriculture)} & \textbf{Agenda 3 (Habitat Preservation)} & \textbf{Agenda 4 (Combatting Corporate Corruption)} & \textbf{Agenda 5 (Removing Toxins from the Environment)} \\
\midrule
Topic 1 & 4883 & 2974 & 148 & 0 & 257 & 716 \\
Topic 2 & 4468 & 1074 & 58 & 2 & 722 & 442 \\
Topic 3 & 8784 & 473 & 23 & 0 & 212 & 177 \\
Topic 4 & 15865 & 187 & 12 & 4 & 102 & 76 \\
\bottomrule
\end{tabular}
\caption{Contingency table of STM-derived \#MAHA topics by MAHA campaign agenda classification.}
\label{tab:s3}
\end{table}

\begin{table}[htbp]
\centering
\begin{tabular}{L{0.06\textwidth} C{0.18\textwidth} C{0.10\textwidth} C{0.10\textwidth} C{0.10\textwidth} C{0.10\textwidth} C{0.10\textwidth}}
\toprule
 & \textbf{No campaign agenda} & \textbf{Agenda 1} & \textbf{Agenda 2} & \textbf{Agenda 3} & \textbf{Agenda 4} & \textbf{Agenda 5} \\
\midrule
Topic 1 & -75.19 & 73.74 & 15.09 & -1.28 & -1.49 & 27.13 \\
Topic 2 & -36.15 & 12.98 & 3.30 & 1.14 & 39.22 & 15.63 \\
Topic 3 & 26.74 & -22.72 & -5.04 & -1.35 & -5.90 & -9.65 \\
Topic 4 & 67.57 & -52.32 & -10.86 & 1.39 & -23.30 & -26.34 \\
\bottomrule
\end{tabular}
\caption{Standardized residuals from the Pearson chi-square test of independence between STM-derived \#MAHA topics and MAHA campaign agenda categories. Absolute residuals greater than 2 indicate notable deviations from expected counts.}
\label{tab:s4}
\end{table}

\newpage
\begin{table}[htbp]
\centering
\begin{tabular}{L{0.06\textwidth} C{0.105\textwidth} C{0.105\textwidth} C{0.10\textwidth} C{0.10\textwidth} C{0.08\textwidth} C{0.13\textwidth} C{0.105\textwidth}}
\toprule
 & \textbf{Facebook} & \textbf{Instagram} & \textbf{Reddit} & \textbf{TikTok} & \textbf{X} & \textbf{Web articles} & \textbf{YouTube} \\
\midrule
Topic 1 & 343 & 125 & 339 & 6565 & 967 & 568 & 71 \\
Topic 2 & 583 & 113 & 244 & 428 & 918 & 4406 & 74 \\
Topic 3 & 656 & 106 & 86 & 555 & 1303 & 6818 & 145 \\
Topic 4 & 126 & 67 & 15 & 15355 & 621 & 49 & 13 \\
\bottomrule
\end{tabular}
\caption{Contingency table of STM-derived \#MAHA topics and online sources.}
\label{tab:s5}
\end{table}

\begin{table}[htbp]
\centering
\begin{tabular}{L{0.06\textwidth} C{0.105\textwidth} C{0.105\textwidth} C{0.10\textwidth} C{0.10\textwidth} C{0.08\textwidth} C{0.13\textwidth} C{0.105\textwidth}}
\toprule
 & \textbf{Facebook} & \textbf{Instagram} & \textbf{Reddit} & \textbf{TikTok} & \textbf{X} & \textbf{Web articles} & \textbf{YouTube} \\
\midrule
Topic 1 & -1.51 & 4.39 & 17.96 & 39.02 & 6.04 & -52.41 & 0.80 \\
Topic 2 & 20.47 & 6.22 & 13.89 & -87.89 & 13.80 & 73.12 & 3.88 \\
Topic 3 & 15.19 & 1.25 & -6.64 & -111.05 & 16.87 & 104.71 & 10.20 \\
Topic 4 & -27.36 & -9.48 & -19.90 & 129.69 & -30.13 & -101.75 & -12.43 \\
\bottomrule
\end{tabular}
\caption{Standardized residuals from the Pearson chi-square test of independence between STM-derived \#MAHA topics and online sources. Absolute residuals greater than 2 indicate notable deviations from expected counts.}
\label{tab:s6}
\end{table}

\newpage
\begin{table}[htbp]
\centering
\begin{tabular}{L{0.10\textwidth} C{0.15\textwidth} C{0.14\textwidth} C{0.13\textwidth} C{0.10\textwidth} C{0.10\textwidth} C{0.10\textwidth} C{0.10\textwidth}}
\toprule
 & \textbf{N} & \textbf{Campaign agenda mention} & \textbf{Interaction} & \textbf{Topic 1} & \textbf{Topic 2} & \textbf{Topic 3} & \textbf{Topic 4} \\ 
 \midrule
 & \textbf{N (\%)} & \textbf{N (\%)} & \multicolumn{5}{c}{\textbf{M (SD)}} \\
\midrule
Facebook & 1710 (4.1\%) & 505 (29.5\%) & 898 (5251) & 0.21 (0.22) & 0.33 (0.23) & 0.36 (0.26) & 0.11 (0.17) \\
Instagram & 412 (1.0\%) & 153 (37.1\%) & 26382 (77513) & 0.30 (0.27) & 0.28 (0.20) & 0.27 (0.22) & 0.16 (0.23) \\
Reddit & 685 (1.6\%) & 110 (16.1\%) & 1693 (6920) & 0.39 (0.20) & 0.34 (0.18) & 0.19 (0.15) & 0.08 (0.09) \\
TikTok & 23045 (55.1\%) & 4018 (17.4\%) & 7389 (109926) & 0.28 (0.30) & 0.06 (0.09) & 0.08 (0.10) & 0.58 (0.33) \\
X & 3809 (9.1\%) & 778 (20.4\%) & 3111 (21909) & 0.24 (0.21) & 0.26 (0.18) & 0.32 (0.24) & 0.18 (0.21) \\
Web articles & 11855 (28.3\%) & 1982 (16.7\%) & 9 (127) & 0.09 (0.13) & 0.37 (0.23) & 0.51 (0.26) & 0.03 (0.05) \\
YouTube & 303 (0.7\%) & 113 (37.3\%) & 27750 (142994) & 0.24 (0.22) & 0.28 (0.18) & 0.37 (0.24) & 0.10 (0.12) \\
\bottomrule
\end{tabular}
\caption{Standardized summary statistics of post count, campaign agenda-aligned post count, interaction metrics provided by NewsWhip Spike, and STM-derived topic prevalence.}
\label{tab:s7}
\end{table}

\newpage
\begin{longtable}{p{0.95\textwidth}}
\caption{AI-assisted coding instructions used for relevance annotation and MAHA campaign agenda identification.}
\label{tab:s8}\\

\toprule
\textbf{Post relevance annotation} 
You are an AI agent that classifies the relevance of social media posts, determining whether a post is relevant to the Make America Healthy Again (MAHA) movement. A post is considered Relevant (code as 1) if it refers to the MAHA/MAGA movement; key individuals such as Donald Trump, Robert F. Kennedy Jr. (RFK), and Biden; key institutions such as congress, CDC, FDA; key issues related to healthcare or wellbeing (e.g., nutrition, carnivore). Otherwise, code it as 0 (not relevant). Output your classification result (i.e., 0, 1) without explanation or additional text. \\

\hline
\textbf{Campaign agenda identification}

\begin{minipage}[t]{0.95\textwidth}
You are an AI agent that classifies social media posts, determining whether a post is relevant to the campaign agenda of the Make America Healthy Again (MAHA) movement. If relevant, assign it to one of the numbered agenda categories below. If not relevant, assign 0. Output your classification result of category numbers only (e.g., 0, 5) without explanation or additional text.

\vspace{0.35em}
\textbf{0 = Not Relevant.}
\begin{itemize}
\item If the post does not relate to any of the agenda topics, classify it as 0.
\end{itemize}

\vspace{0.35em}
\textbf{1 = Chronic Disease Epidemic.}
\begin{itemize}
\item Classify as 1 if the post discusses diet, nutrition, preventative healthcare, lifestyle, exercise, obesity, diabetes, or chronic health conditions.
\item Common indicators include: chronic, illness, disease, obesity, diabetes, metabolism, hypertension, cardiovascular, cancer, stroke, exercise, diet, anxiety, depression.
\item If a post is about diet/healthy eating/food access without toxins, use 1.
\end{itemize}

\vspace{0.35em}
\textbf{2 = Regenerative Agriculture.}
\begin{itemize}
\item Classify the post as 2 if it talks about farming practices, soil health, sustainable agriculture, regenerative methods, or biodiversity in agriculture.
\item Common indicators include: agriculture, farm, soil, regenerative, covercrop, crop, no-till, compost, biodiversity, permaculture, pollinator.
\item If pesticides or foods are discussed as part of farming practices, use 2.
\end{itemize}

\vspace{0.35em}
\textbf{3 = Habitat Preservation.}
\begin{itemize}
\item Classify the post as 3 if it focuses on protecting or restoring natural habitats, conserving ecosystems, preserving forests, wetlands, wildlife, or endangered species.
\item Common indicators include: habitat, preserve, conserve, ecosystem, restore, forest, wildlife, wetland, river, endangered, species.
\item If conservation or restoration is discussed and not centered on pollution, use 3.
\end{itemize}

\vspace{0.35em}
\textbf{4 = Combatting Corporate Corruption.}
\begin{itemize}
\item Classify the post as 4 if it highlights corporate influence, lobbying, lack of transparency, accountability issues, or corruption related to health and environment.
\item Common indicators include: corporate, pharmaceutical, pharma, monsanto, bayer, pfizer, lobby, reform, transparent, accountable, regulate, dark money, campaign finance.
\end{itemize}

\vspace{0.35em}
\textbf{5 = Removing Toxins from the Environment.}
\begin{itemize}
\item Classify the post as 5 if it addresses chemicals, pollutants, contamination in air, water, or food, or toxin-free initiatives.
\item Common indicators include: toxic, toxin, clean, fluorid, pollute, contaminate, pfas, chemical, hazard, lead, mercury, arsenic, safe, glyphosate, herbicide, pesticide.
\item If a post is about toxins, chemicals, pollution, ``clean food'' framed as toxin removal, or pesticides framed as pollution exposure, use 5.
\end{itemize}
\end{minipage}
\\
\hline

\textbf{Campaign agenda full text}

\begin{minipage}[t]{0.95\textwidth}
\vspace{0.35em}
\textbf{1 = Chronic Disease Epidemic.}
\begin{itemize}
\item ``We advocate for a comprehensive, national strategy to address the root causes of America's chronic disease crisis---poor diets, environmental toxins, and inadequate healthcare systems. By promoting preventative healthcare and clean food initiatives, we aim to empower individuals and communities with access to healthier lifestyles.''
\end{itemize}

\vspace{0.35em}
\textbf{2 = Regenerative Agriculture.}
\begin{itemize}
\item ``We are leading the charge for sustainable farming practices that rebuild our soil, reduce chemical reliance, and promote biodiversity. Regenerative agriculture is the key to healthier food systems and a major player in the fight to Make America Healthy Again.''
\end{itemize}

\vspace{0.35em}
\textbf{3 = Habitat Preservation.}
\begin{itemize}
\item ``Environmental preservation is integral to public health. We fight for stronger protections of America's natural ecosystems and promote policies that restore and sustain these precious resources for generations to come.''
\end{itemize}

\vspace{0.35em}
\textbf{4 = Combatting Corporate Corruption.}
\begin{itemize}
\item ``We are a watchdog for the public interest, exposing the corporate stranglehold on the very government agencies meant to protect our health and environment. By pushing for transparency and accountability, we work to restore trust in our nation's institutions.''
\end{itemize}

\vspace{0.35em}
\textbf{5 = Removing Toxins from the Environment.}
\begin{itemize}
\item ``Clean air, water, and food are non-negotiables. We promote candidates and policies that remove harmful chemicals and pollutants to ensure every American lives in a toxin-free environment.''
\end{itemize}
\end{minipage}
\\
\bottomrule
\end{longtable}

\end{document}